\documentclass[nologo,11pt,a4paper]{ETHpaper}
\pdfoutput=0

\usepackage{amssymb,amsmath}                   % Mathematical Symbols
\usepackage{graphicx, epsfig}                          % Figures
\usepackage[english]{babel}                    % Spell-checking
\usepackage[]{hyperref} % Hyperlinks
\usepackage[square,numbers,sort&compress]{natbib}

\usepackage[utf8]{inputenc}

\title{A $k$-shell decomposition method for weighted networks}

\author{Antonios Garas$^{1}$, Frank Schweitzer$^{1}$ and Shlomo Havlin$^{2}$}

\address{$^{1}$Chair of Systems Design, ETH Zurich, Weinbergstrasse 58
  CH-8092 Zurich, Switzerland\\ $^{2}$ Minerva Center and Department
  of Physics, Bar-Ilan University, 52900 Ramat Gan, Israel}

\authoralternative{A. Garas, F. Schweitzer and S. Havlin}

\reference{\textsl{New J. Phys.} \textbf{14} (2012) 083030}

\www{\url{http://www.sg.ethz.ch}}

\newcommand{\mean}[1]{\left\langle #1 \right\rangle}

\begin{document}

\maketitle

\begin{abstract}
  We present a generalized method for calculating the k-shell
  structure of weighted networks. The method takes into account both
  the weight and the degree of a network, in such a way that in the
  absence of weights we resume the shell structure obtained by the
  classic k-shell decomposition. In the presence of weights, we show
  that the method is able to partition the network in a more refined
  way, without the need of any arbitrary threshold on the weight
  values. Furthermore, by simulating spreading processes using the
  susceptible-infectious-recovered model in four different weighted
  real-world networks, we show that the weighted k-shell decomposition
  method ranks the nodes more accurately, by placing nodes with higher
  spreading potential into shells closer to the core. In addition, we
  demonstrate our new method on a real economic network and show that
  the core calculated using the weighted k-shell method is more
  meaningful from an economic perspective when compared with the
  unweighted one.
\end{abstract}

%\pacs{89.75.-k, 89.75.Fb, 05.10.-a}

\section{Introduction}

The continuously growing attention in complex network science resulted
over the past years in novel ways of analysis for a great number of
complex systems in various scientific fields
\cite{Albert2002,Dorogovtsev2003,Caldarelli2007,Barrat2008a,Cohen2010,Newman2010a,Jackson2010a}.
The fundamental view of this interdisciplinary approach is that large
complex systems can be described as complex networks (or graphs under the
mathematics terminology) where the nodes (or vertices) represent the
system's interacting elements and the links (or edges) represent their
interactions. This unified view was used in the analysis of social
\cite{Jackson2010a,Snijders2006,Borgatti2006}, biological
\cite{Milo2002,Alon2003,Khanin2006,Gallos2012}, physiological
\cite{Bashan2012}, technological \cite{Reed2009,Johansson2010}, climate
\cite{Tsonis2008,Donges2009,Gozolchiani2011}, economic
\cite{Schweitzer2009b,Garas2010,Vitali2011,Harmon2010}, and financial
systems \cite{Bonanno2003,Garas2008a}. In combination with the
technological advances that made enormously detailed data available, we
are now able to understand and model the evolution of dynamical
processes, like epidemic outbreaks and information spreading
\cite{Daley1965,Colizza2006,Castellano2007,Yang2008a,Yang2008b}.

Even the earliest empirical works in this field made clear to researchers
that the topology of a network affects its properties. For example,
networks with broad degree distributions are more robust to random
failures, but are fragile under intentional attacks
\cite{Albert2000b,Cohen2000,Callaway2000,Cohen2001a,Gallos2005}.
Nowadays, there is a growing body of literature trying to understand
global properties of a network by focusing on properties of individual
nodes, and their connectivity patterns \cite{Park2007}. Of course the
role of individual nodes has a profound relation to the evolution of any
dynamical process, and to the evolution of the network itself. For
example, very popular individuals in a social network (i.e. individuals
with a large number of connections) usually attract more attention and
increase even more their connectivity. While it is clear that such
processes affect the evolution of the network topology, we can imagine
that such individuals could assume key roles in the case of disease
spreading etc.

It is clear that questions like {\it "Who are the most important nodes in
  the network?"} are natural to ask. Such questions can be addressed
using centrality measures, which are the most frequently used measures
when it comes to quantitative network analysis. However, there is a
variety of centrality measures aiming to address the question of node
"importance". For example there is the {\it degree centrality} (or just
the degree of a node, i.e. the number of its links), the {\it eigenvector
  centrality} \cite{Bonacich1987a}, the {\it betweenness centrality}
\cite{Freeman1977}, the {\it closeness centrality} \cite{Newman2005b},
etc. In this work we focus on a centrality measure based on the notion of
$k$-cores which is a fundamental concept in Graph
Theory~\cite{Bollobas1984} when it comes to ranking the centrality of
nodes in a complex network. Such ranking was applied in many real
networks~\cite{Seidman1983, Bader2003, Wuchty2005, Dorogovtsev2006,
  Carmi2007, Alvarez-Hamelin2008, Shao2009, Garas2010, Kitsak2010}
allowing a thorough investigation of their structure, while highlighting
the role of various topology-dependent processes.

One major limitation of most centrality measures, including the $k$-core
decomposition method, is their design to work on unweighted
graphs. However, in practice, real networks are weighted, and their
weights describe important and well defined properties of the underlying
systems. In a weighted network, nodes have (at least) two properties that
can characterise them, their degree and their weight. However, since
weights are properties of the network's links, the node's weight is
calculated as the sum over all link weights passing through a particular
node. These two properties, even though in some cases are correlated, are
in general independent. As a result, nodes with high degree can have
small weight (i.e. they have many connections with other nodes but the
links of these connections have small weights), while there could also be
nodes with small degree and high weight. Situations where the weights
play important role, occur for example in economic or trade networks. In
such networks the weights are related to some measured property (like
trade flow, capital flow etc.), and in many cases one wishes to focus on
nodes with high weights that are (usually) the most important
players. Thus, in such systems the presence of nodes with high degree and
relatively small weights may influence the results obtained by methods
that are based only on the degree.  In such cases two main approaches
have been used, with both having their own drawbacks. Under the first
approach one completely neglects the weights and performs the analysis on
the unweighted network, but doing so one chooses to neglect an important
property of the network. The second approach would be to consider only
links with weights above some - (usually) arbitrary chosen - threshold
value and filter out the rest. The drawback of this approach is the
selection of a proper cut-off value, which may remove important high
degree nodes with links of low weights (below the threshold) and as we
will discuss later, this could have significant impact on the
results. Additionally, by neglecting links below a threshold, the network
becomes sparser with some nodes getting disconnected and not considered
by the applied method afterwards.

Here we aim to overcome these failures by introducing a generalized
method for calculating the $k$-shell structure of weighted networks. The
paper is organized in the following way: first we discuss the standard
$k$-shell decomposition method, and right after we introduce our
generalized version. Next, we apply both methods on real networks and we
present their results. Subsequently we compare in more detail the
performance of both methods in ranking nodes according to their
importance when it comes to spreading processes, and at the end we
summarize our conclusions.

\begin{figure}
  \begin{center}
    \includegraphics[scale=0.75]{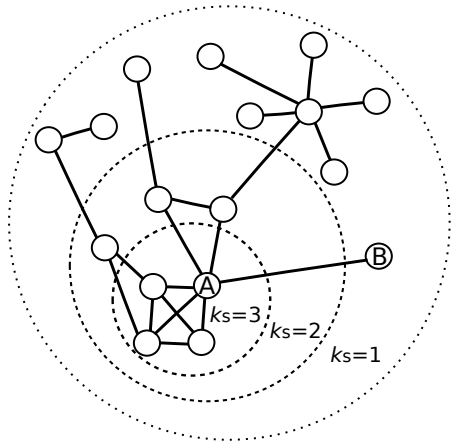}
  \end{center}
  \caption{Illustration of the layered structure of a network, obtained
    using the $k$-shell decomposition method. The nodes between the two
    outer rings include nodes of shell 1 ($k_s=1$), while the nodes
    between the two inner rings compose shell 2 ($k_s=2$). The nodes
    within the central ring constitute the core, in this case $k_s=3$.}
  \label{fig:gsh.example}
\end{figure}

% \section{Methods}
% \label{sec:gsh.method}

\section{The unweighted $k$-shell decomposition method} 

The k-core/k-shell decomposition method partitions a network into
sub-structures that are directly linked to centrality
\cite{Batagelj2010}. This method assigns an integer index, $k_{s}$, to
each node that is representative of the location of the node in the
network, according to its connectivity patterns. Nodes with low/high
values of $k_{s}$ are located to the periphery/center of the
network. This way, the network is described by a layered structure
(similar to the structure of an onion), revealing the full hierarchy of
its nodes. The innermost nodes belong to the structure called core or
"nucleus" of the network, while the remaining nodes are placed into more
external layers ($k$-shells).

A more detailed description of how a network is divided into this
$k$-shell structure is the following (see
Fig~\ref{fig:gsh.example}). First one removes recursively from the
network all nodes with degree $k=1$, and we assign the integer value
$k_{s}=1$ to them. This procedure is repeated iteratively until there are
only nodes with degree $k\geq 2$ left in the network. Subsequently, one
removes all nodes with degree $k=2$ and assign to them the integer value
$k_{s}=2$. Again, this procedure is repeated iteratively until there are
only with nodes with degree $k\geq 3$ left in the network, and so
on. This routine is applied until all nodes of the network have been
assigned to one of the $k$-shells. This is how the original $k$-shell
decomposition method works, which, as described above, does not consider
at all the weights of the links; therefore, from now on we will call it
unweighted $k$-shell decomposition method ($U_{k-{\rm shell}}$).

\section{The weighted $k$-shell decomposition method}

Here we propose a generalization of the $k$-shell decomposition method,
that we call weighted $k$-shell decomposition method ($W_{k-{\rm
    shell}}$). This method applies the same pruning routine that was
described earlier, but it is based on an alternative measure for the node
degree. This measure considers both the degree of a node and the weights
of its links, and we assign for each node a {\it weighted degree},
$k'$. The weighted degree of a node $i$ is defined as
\begin{equation}
  k'_{i}=\left[k_{i}^{\alpha} \left( \sum_{j}^{k_{i}}{w_{ij}}
    \right)^{\beta} \right]^{\frac{1}{\alpha+\beta}},
  \label{eq:gsh.kprime}
\end{equation}
% \begin{equation}
%   k'_{i}=\left[k_{i} \sum_{j}^{k_{i}}{w_{ij}}\right]^{1/2},
%   \label{eq:gsh.kprime}
% \end{equation}
where $k_{i}$ is the degree of node $i$, and $\sum_{j}^{k_{i}}{w_{j}}$ is
the sum over all its link weights. In the present study we discuss only
the case where $\alpha=\beta=1$, which treats the weight and the degree
equally. The full exploration of the parameter space is outside our
scope, and is left for future work. Therefore, for what follows
$k'_{i}=\sqrt{k_{i} \sum_{j}^{k_{i}}{w_{ij}}}$.

Using the above approach in the case of unweighted networks, where
$w_{ij}=1$, the weighted degree is equivalent to the node degree
($k'\equiv k$), and we resume the same network partitioning as with the
$U_{k-{\rm shell}}$ decomposition method. However, in order that a
typical weighted link will be regarded as of unit weight before we
calculate $k'$ using Eq.~\ref{eq:gsh.kprime} we perform the following
steps. First, we normalize all the weights with their mean value $\mean
w$, next we divide the resulting weights with their minimum value, and we
discretize them by rounding to the closest integer; this way the minimum
link weight is equal to one~\footnote{We also tested the effect of the
  normalization by dividing with the minimum weight, and the results we
  obtained in terms of node positioning with or without the normalization
  were similar.}.

In Fig.~\ref{fig:gsh.example} we illustrate schematically the layered
structure obtained by applying the $U_{k-{\rm shell}}$ decomposition
method in a graph. In order to highlight the weaknesses of the unweighted
method, let us suppose that the network is weighted. For simplicity we
assume that all link weights are equal to one, except for the weight of
the link between nodes $A$ and $B$, which is $w_{AB}=3$.  As illustrated
in Fig.~\ref{fig:gsh.example}, the node $B$ is located at the periphery
of the network, even though it is strongly connected to one of the core
nodes. In real networks such a strong link (3 times the capacity of other
links) means that this particular node is of more importance for the
core, but this is not depicted in the layered structured calculated by
the classical unweighted approach, since this node will be placed in the
outermost shell ($k_{s}=1$). However, if we apply the
%$k$-shell method using the weighted degree (which we call 
$W_{k-{\rm shell}}$ decomposition method, then node $B$ is assigned to
$k_{s}=2$ that is one shell away from the core of the network,
highlighting its actual importance.

\section{Application to real networks}

% \subsection{Data}
% \label{sec:gsh.data}

\begin{figure}[tb]
  \begin{center}
    \includegraphics[scale=0.42]{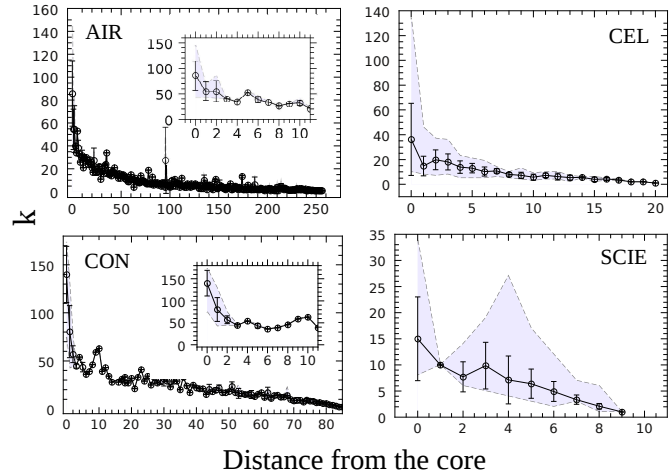}
  \end{center}
  \caption{Average degree of all nodes in each shell, obtained using the
    $W_{k-{\rm shell}}$ decomposition method. The shaded area highlights
    the full range of the degree values in each shell. The shells are
    ranked according to their distance from the core, and the error bars
    are showing the standard deviation. Insets: zoom to distances closer
    to the core for networks with large number of shells}.
  \label{fig:gsh.All}
\end{figure}

In order to compare between the results obtained from the $U_{k-{\rm
    shell}}$ and the $W_{k-{\rm shell}}$ decomposition method, we used as
case studies the following four real networks:

\begin{enumerate}
\item {\it Corporate Ownership Network} (CON). This is an economic
  network linking 206 different countries. It is
  constructed~\cite{Garas2010} using the 616000 direct or indirect
  subsidiaries of the 4000 world corporations with the highest turnover,
  based on the 2007 version of the ORBIS database obtained from the {\it
    Bureau van Dijk Electronic Publishing} (BvDEP)~\footnote{Bureau van
    Dijk Electronic Publishing (BvDEP) http://www.bvdep.com/}. The
  network is weighted, and it's weights represent the business ties among
  countries~\cite{Garas2010}.
 
\item {\it The collaboration network of scientist working in network
    science} (SCIE). This network contains the Co-authorship relations of
  scientists working on network theory and experiment, as compiled by
  M. Newman~\cite{Newman2006}. The network is weighted, and it's weights
  are assigned as described in~\cite{Newman2001b}.

\item {\it The neural network of the nematode C. Elegans} (CEL). This
  network was compiled by D. Watts and S. Strogatz~\cite{Watts1998}
  using the original experimental data by White et
  al~\cite{White1986}. It is a weighted representation of the neural
  network of C. Elegans.

\item {\it The U.S. Air transportation network} (AIR). This is a weighted
  network obtained by considering the 500 US airports with the largest
  amount of traffic from publicly available
  data~\cite{Colizza2007}. Nodes represent US airports and edges
  represent air travel connections among them. It reports the anonymized
  list of connected pairs of nodes and the weight associated to the edge,
  expressed in terms of number of available seats on the given connection
  on a yearly basis.
\end{enumerate}

In Table~\ref{tab:gsh.NetStats} we provide some detailed statistical
properties of the above networks. For our analysis, if not stated
otherwise, when we talk about the network we refer to the largest
connected component (LCC), and whenever we discuss network properties
these are calculated from the LCC.

In Table~\ref{tab:gsh.NetCoreComp} we compare the network hierarchies
obtained by applying the $U_{k-{\rm shell}}$ and the $W_{k-{\rm shell}}$
decomposition method. We observe that the $W_{k-{\rm shell}}$ method
yields a more refined partitioning (larger number of $k$-shells) of the
networks. This means that by applying this method we obtain more detailed
information about the networks' internal structure, and is similar to
using a high resolution microscope to observe small size structures of a
larger system.

\begin{table}[htbp]
  \caption{Statistical properties of the networks used in our
    analysis. Here $N_N$ is the number of nodes, $N_E$ is the number of
    edges, $\mean k$ is the average degree of the network nodes, $d$ the
    diameter, $C$ the clustering coefficient~\cite{Watts1998}, and $B$
    the network's betweenness~\cite{Freeman1977,Freeman1979}. If the
    original network is disconnected, we only consider it's largest
    connected component.} 
\begin{center}
%\begin{ruledtabular}
  \begin{tabular}{lllllll}
    %\hline
%    \hline
    Network  & $N_N$ & $N_E$ & $\mean k$& $d$	& $C$     & $B$   \\ \hline
    CON      & 206   & 2886  & 28.0  	& 4  	& 0.38    & 94.6  \\ %\hline
    SCIE     & 379   & 914   & 4.82  	& 17 	& 0.43    & 952.9 \\ %\hline
    CEL      & 297   & 2345  & 15.8  	& 5  	& 0.18    & 215.4 \\ %\hline
    AIR      & 500   & 2980  & 11.92    & 7     & 0.35    & 496.7  %\hline
    %\hline
  \end{tabular}
%\end{ruledtabular}
\label{tab:gsh.NetStats}
\end{center}
\end{table}

\begin{table}[htbp]
  \caption{Comparison of the network hierarchies obtained by  the
    $U_{k-{\rm shell}}$ and $W_{k-{\rm shell}}$ decomposition method.
    Here $s^{U}$ and $s^{W}$ is the total number of $k$-shells, while
    $n_{c}^{U}$ and $n_{c}^{W}$ is the total number of nodes in the cores
    obtained using the $U_{k-{\rm shell}}$  and the $W_{k-{\rm shell}}$
    respectively. $N_{C}$ is the number of common nodes in both cores,
    $N_{UW}$, is the fraction of nodes that belong to the core obtained
    by the $U_{k-{\rm shell}}$ that also belong to the core obtained by
    the $W_{k-{\rm shell}}$, and $N_{WU}$ is the fraction of nodes of the
    core obtained by the $W_{k-{\rm shell}}$ that also belong to the core
    obtained by the $U_{k-{\rm shell}}$.}
\begin{center}
%\begin{ruledtabular}
  \begin{tabular}{lllllccc}
    %\hline
    %\hline
    Network& $s^{U}$ & $s^{W}$& $n_{c}^{U}$& $n_{c}^{W}$	&
    $N_{C}$ & $N_{UW}$    & $N_{WU}$ 	\\ \hline
    CON    & 28	&  87    & 41	 & 11	&  11    & 0.27& 1	\\ %\hline
    SCIE   & 8 	&  10    & 9	 & 13	&  9     & 1   & 0.69	\\ %\hline
    CEL    & 10	&  21    & 119   & 26	&  26    & 0.22& 1	\\ %\hline
    AIR    & 29 &  257   & 35    & 31   &  28    & 0.8 & 0.9     %\hline
    %\hline
  \end{tabular}
%\end{ruledtabular}
\label{tab:gsh.NetCoreComp}
\end{center}
\end{table}

Furthermore, for three out of the four studied networks the core obtained
with the $W_{k-{\rm shell}}$ contains smaller number of nodes, while
these nodes are almost entirely part of the core obtained by the
$U_{k-{\rm shell}}$. This means that the weighted method in most cases is
able to split further the cores obtained by the unweighted method and to
identify which are the {\it most central of the central nodes}.

In Fig.~\ref{fig:gsh.All} we plot the degrees of the nodes according to
the $k$-shell they belong (expressed as the distance from the core of the
network). The node ranking is obtained using the $W_{k-{\rm shell}}$
method for all the four different networks described above. As shown in
Fig.~\ref{fig:gsh.All}, the degree is highly (and non linearly)
correlated with the position of the node in the $k$-shell structure, but
there are particular cases where the trend is not monotonous. This means
that there are nodes with high degree that may not be as central to the
network as one would expect; this is in line with our discussion for the
example network of Fig.~\ref{fig:gsh.example}.

\subsection{A detailed example: analysis of the core of an economic
  network}
\label{sec:gsh.ex.econet}

\begin{figure*}[htbp]
  \begin{center}
    \includegraphics[scale=0.6,angle=-90]{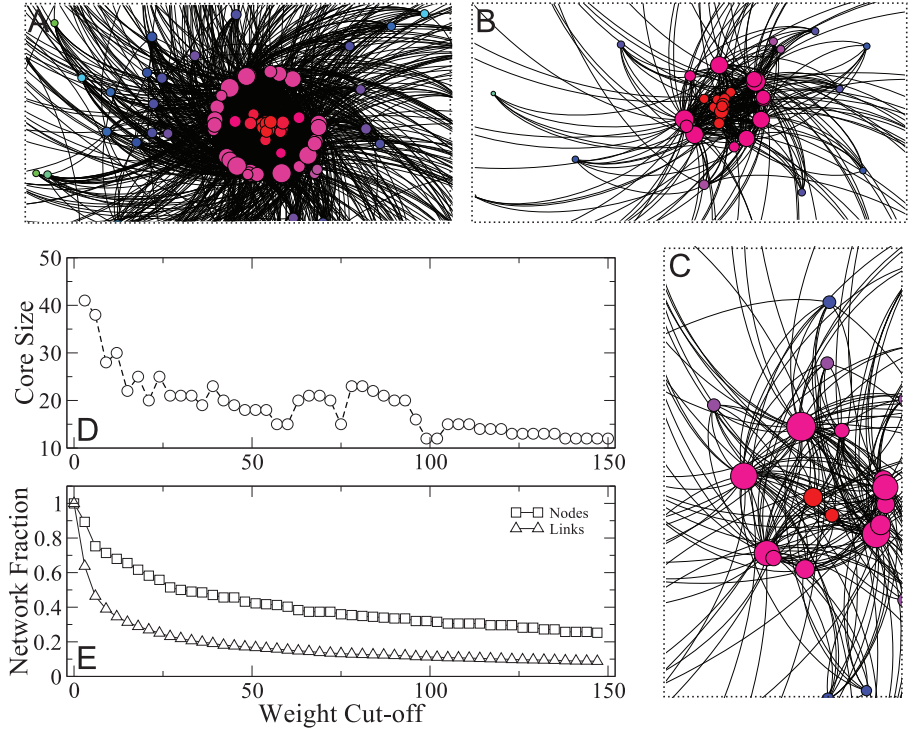}
  \end{center}
  \caption{Changes in the CON network structure when using different
    weight cut-off values $w_{c}$. Panels A), B), and C) show the network
    snapshots around the central region for $w_{c}=3$, $w_{c}=75$, and
    $w_{c}=150$ respectively. The size of the nodes is proportional to
    their degree. D) Evolution of the core size as a function of $w_{c}$
    (After Garas et al \cite{Garas2010}). E) Fraction of nodes and links
    of the original network that remain for different $w_{c}$ values.}
  \label{fig:gsh.CON.CoreSize}
\end{figure*}

Next we compare the core of the $U_{k-{\rm shell}}$ and the $W_{k-{\rm
    shell}}$ decomposition methods applied on the global Corporate
Ownership Network (CON) studied in~\cite{Garas2010}. The CON connects
206 countries around the globe, using as links the ownership relations
within large companies.  If companies listed in country A have subsidiary
corporations in country B, there is a link connecting these two countries
directed from country A to country B. The weight of the link, $w_{AB}$,
equals the number of the subsidiary corporations in country B controlled
by companies of country A.

Using the $U_{k-{\rm shell}}$ decomposition method, as shown in
Table~\ref{tab:gsh.NetCoreComp} and Fig.~\ref{fig:gsh.CON.CoreSize}, we
identify a core of 41 countries. However we expect that in the current
state of the global economy a smaller set of countries are the major
players (G8, G20, etc). In order to reduce the size of the core, and to
highlight which are the potentially more important nodes of this network
by using the classic $k$-shell decomposition method, a cut-off value of
$w_{c}=100$ was assumed in Garas et al~\cite{Garas2010}. It was shown
that the remaining network after filtering the links with $w_{c}<100$
contains only 66 out of the original 206 nodes. However, a core formed by
the following 12 countries: United States of America (US), United Kingdom
(GB), France (FR), Germany (DE), Netherlands (NL), Japan (JP), Sweden
(SE), Italy (IT), Switzerland (CH), Spain (ES), Belgium (BE), and
Luxembourg (LU) was identified. In Fig.~\ref{fig:gsh.CON.CoreSize} the
evolution of the core and network size of the CON is shown, as a function
of the weight cut-off value $w_{c}$.

Using the $W_{k-{\rm shell}}$ decomposition method we obtain the layered
structure of the network including all the 206 nodes, without using any
arbitrary cut-off parameter. The core of the network obtained with this
method consists of the following 11 counties: United States of America
(US), United Kingdom (GB), France (FR), Germany (DE), Netherlands (NL),
Japan (JP), Canada (CA), Italy (IT), Switzerland (CH), Spain (ES), and
Belgium (BE). Comparing these two cores we find a striking
similarity. The only two differences are the presence of Canada (CA) in
the core calculated using our new weighted $k$-shell approach while
Sweden (SE) and Luxembourg (LU) have moved to the second innermost
layer. These differences can be well understood, considering that CA is a
major economy, it is part of G7, and all the other six members of G7 are
already part of the core. Furthermore, CA outperforms SE and LU in terms
of population and other macroeconomic indicators, such as total
import/exports and GDP.  It is thus natural to conclude that the core
obtained using the $W_{k-{\rm shell}}$ decomposition method is more
meaningful from an economics perspective, since it groups together some
of the largest (developed) global economies.

\section{Dynamics: Shell positioning and spreading potential}
\label{sec:gsh.shvsspr}

Recently models like the Susceptible-Infectious-Recovered (SIR)
model~\cite{Anderson1991} have been used extensively in network research
in order to explore epidemic
spreading~\cite{Anderson1991,Hethcote2000,Newman2002,Colizza2006},
economic crisis spreading~\cite{Garas2010} as well as information and
rumor spreading~\cite{Daley1965,Castellano2007} in social processes.
However, in such processes the topology of the network is not the only
thing that matters; the position of the node where the spreading begins
plays an important role as well. In the resent work of Kitsak et
al~\cite{Kitsak2010} it was shown that the spreading power of a node
cannot be predicted solely based on its degree. A better measure is its
actual position in the network, as it is described by the $k$-shell where
it belongs.

Using this perspective, it is reasonable to assume that a $k$-shell
partitioning method provides us with a more accurate node ranking for
representing the nodes' spreading power. Additionally, since the
individual nodes are grouped in $k$-shells, it is reasonable to assume
that every $k$-shell should contain nodes with similar spreading
power. In what follows we will use these assumptions to evaluate and
compare the performance of the $U_{k-{\rm shell}}$ and $W_{k-{\rm
    shell}}$ decomposition methods.

We modeled spreading process by applying the SIR model on all the
networks described above. However, since we are interested in the weights
of the network, we used a version of the SIR model which takes into
account the weight of the links that mediate the spreading. This model
was originally introduced to simulate the spreading of an economic
crisis~\cite{Garas2010}; for this model the probability of infection is
different for every link and is calculated by
\begin{equation}
 p_{ij}\propto m\cdot w_{ij}/\tilde{w}_{j},
\label{eq:gsh.wp}
\end{equation}
where $w_{ij}$ is the weight of the link that connects the origin node
$i$ with the destination node $j$, and $\tilde{w}_{j}$ is the total
weight ($\tilde{w}_{j}=\sum_{i}w_{ij}$) of the destination node $j$. The
factor $m$ is a free amplification parameter that can determine for
example the severity of a crisis, how infectious a virus is, the
importance of a rumor etc. In what follows we will call this model {\it
  Weighted SIR} (W-SIR).

\begin{figure}[t]
  \begin{center}
    \includegraphics[scale=0.32]{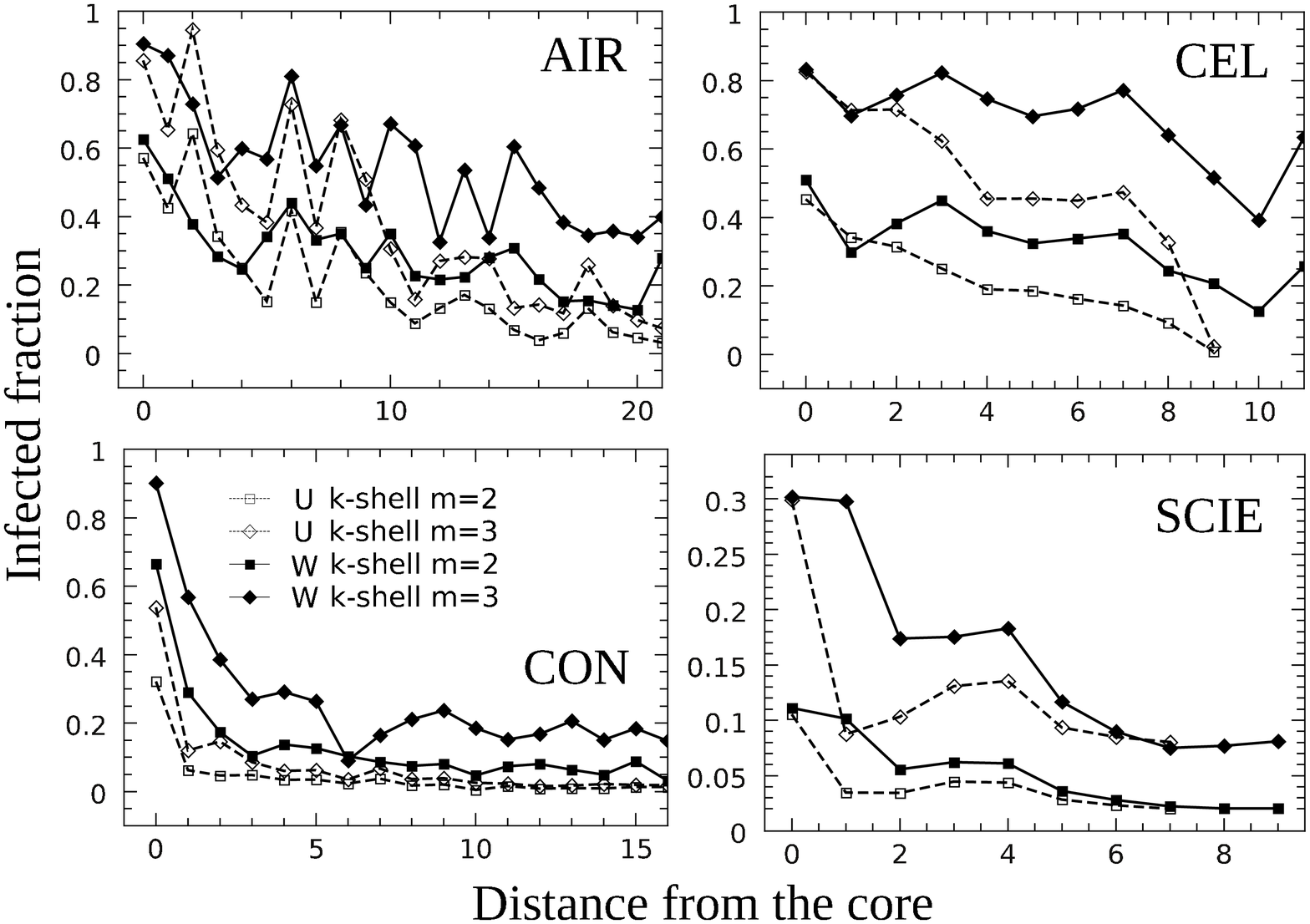}
  \end{center}
 \caption{Average infected fraction of a $k$-shell versus the shell's
   distance from the core of the network.}
 \label{fig:gsh.WSIR-All}
\end{figure}

\begin{figure}[t]
  \begin{center}
    \includegraphics[scale=0.32]{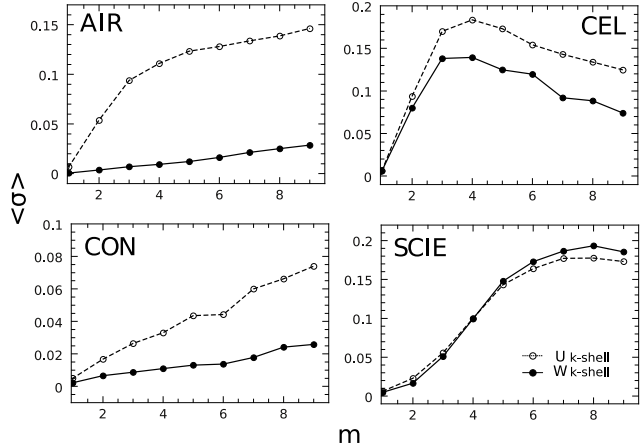}
  \end{center}
 \caption{Average value of the spreading potential of nodes within a
   $k$-shell over all shells, $\mean\sigma$, versus $m$.}
 \label{fig:gsh.WSIR-SD}
\end{figure}

The modeling procedure of the W-SIR is the following. Initially we assign
all nodes to be susceptible (S) to an infection. Next, one node, $i$, is
chosen and is assumed to be infected (I). This node will infect all its
neighboring nodes with probability $p_{ij}$ during the first time
step. This causes all infected nodes to switch their status from S to I,
while the node that initiated this process changes to the recovered state
(R), and can no longer infect other nodes or become infected. At every
consecutive time step the process is repeated, and all the infected nodes
are trying to infect their susceptible (S) neighbors in the network. The
process lasts until there are no infected nodes left in the network.

For each individual node we performed 100 realizations of the W-SIR
model, and we calculated the average infected fraction of the network for
different values of $m\in[0,10]$. This fraction is used as score in order
to rank the nodes according to their spreading potential. We restricted
ourselves to values of $m$ in this interval, as for much larger $m$
values the role of individual nodes is no longer important, and an
epidemic outbreak emerges no matter where the infection starts. Next, we
partitioned the network using the $U_{k-{\rm shell}}$ and the $W_{k-{\rm
    shell}}$ decomposition methods, and ranked the obtained $k$-shells
according to their distance from the core.  By calculating the average
infected fraction that results from an epidemic starting separately from
all nodes of every individual $k$-shell, we estimated the shell's
spreading potential.

In Fig.~\ref{fig:gsh.WSIR-All} we study how the average infected fraction
changes versus the distance of each $k$-shell from the core of the
network for both methods. We find that, in general, the central
$k$-shells obtained by the $W_{k-{\rm shell}}$ method are more able to
initiate a severe outbreak in comparison to the central $k$-shells
obtained using the $U_{k-{\rm shell}}$ method. This result is robust for
all networks used in this study, and for different values of the
parameter $m$. The above finding means that the $W_{k-{\rm shell}}$
decomposition method positions the nodes with the higher average
spreading potential in shells closer to the core.

Next, we tested how homogeneous are the obtained $k$-shells with respect
to the spreading potential of their containing nodes. In order to do so,
we calculated the standard deviation, $\sigma$, of a node's infected
fraction (spreading potential) for every $k$-shell for a given value of
the parameter $m$. Next we calculated the average value over all the
shells, $\mean\sigma$, and we plot it versus $m$
(Fig.~\ref{fig:gsh.WSIR-SD}). We find that the average standard deviation
of the spreading potential using W-SIR is always lower when we partition
the network using the $W_{k-{\rm shell}}$ method, with respect to
partitioning using the $U_{k-{\rm shell}}$ method. This means that the
$W_{k-{\rm shell}}$ method gives more \textit{homogeneous $k$-shells},
where all nodes in the shell have similar importance for the dynamical
process in question.

\begin{figure}[t]
  \begin{center}
    \includegraphics[scale=0.62]{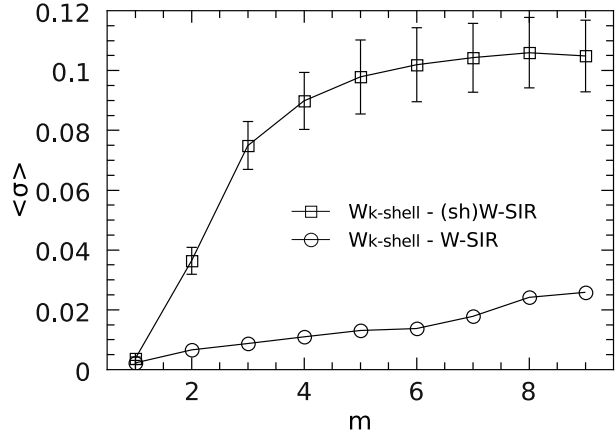}
  \end{center}
 \caption{Comparison of $\mean\sigma$ versus $m$ for two different
   configurations of the CON. $W_{k-{\rm shell}}$ - W-SIR is the original
   case (also shown in Fig.~\ref{fig:gsh.WSIR-SD}) where the nodes'
   spreading potential is obtained by applying the W-SIR in the original
   network. $W_{k-{\rm shell}}$ - (Sh)W-SIR is a case where we calculated
   the nodes' spreading potential by applying the W-SIR on the 10
   realizations of the CON with shuffled weights.}
 \label{fig:gsh.CON.Sh}
\end{figure}

As a final step, and given that the $W_{k-{\rm shell}}$ method performs
better in positioning the nodes according to their W-SIR spreading
potential in weighted graphs, it is interesting to further explore the
role of the weights in this process.  To do so, we created 10
realizations of the CON network with shuffled weights, and we performed
100 runs of the W-SIR model on every one of these 10 networks. Next, we
calculated the average spreading potential per $k$-shell using the
infected fraction obtained by the implementation of W-SIR on the network
with shuffled weights.  As shown in Fig.~\ref{fig:gsh.CON.Sh}, in the
shuffled case the $k$-shells are becoming significantly more
inhomogeneous, and their $\mean\sigma$ is always larger that the
$\mean\sigma$ obtained by the original, unshuffled network.  This
procedure highlights the role of the weights in the process, since in the
case where the weights do not to play any role these two curves should
collapse into one.

\section{Conclusion}

In summary, we presented a generalized $k$-shell decomposition method
($W_{k-{\rm shell}}$) that considers the link weights of networks,
without applying any arbitrary cut-off threshold on their value. The
method resumes the same shell structure obtained by the classic $k$-shell
decomposition in the absence of weights, but when weights are present, it
is able to partition the network in a more refined way. In it's general
formulation, our method allows us to vary the importance assigned to
either the node weights or the node degree, by adjusting the exponents
$\alpha$ and $\beta$ of Eq. \ref{eq:gsh.kprime}. Whilst in the current
work we did not fully explore the parameter space, we would like to
stress that this additional flexibility provides a more accurate ranking
for various applications. Here, using $\alpha=\beta=1$ we showed that the
partitioning obtained by the $W_{k-{\rm shell}}$ method is particularly
meaningful in terms of the spreading potential of the nodes. We
demonstrated the weighted version of the SIR model in four different
networks, and showed that nodes with higher spreading potential were
positioned in the core or in shells closer to the core, better in
comparison with the $U_{k-{\rm shell}}$ method.

\section{Acknowledgement}

S.H. wishes to thank the European EPIWORK and LINC projects, the Israel
Science Foundation, ONR, DFG, and DTRA for financial
support. A.G. acknowledges financial support from the Swiss National
Science Foundation (Project 100014 126865).

%\newpage

\section*{References}
\bibliographystyle{unsrt}

%\bibliographystyle{sg-bibstyle}
%\bibliography{k-shell}   % name your BibTeX data base

\end{document}